\documentstyle[12pt,epsfig]{article}
\newcommand{\lsim}{\raisebox{-0.07cm}{$\, \stackrel{<}{{\scriptstyle
\sim}}\, $}}
\newcommand{\gsim}{\raisebox{-0.07cm}{$\, \stackrel{>}{{\scriptstyle
\sim}}\, $}}
\begin{document}
\thispagestyle{empty}

\phantom{X}
\vspace{2.5cm}

\begin{center}
{\LARGE\bf QCD Corrections to the}

\vspace*{2mm}
{\LARGE\bf Polarized Drell-Yan Process}

\vspace{1cm}
{\large T.~Gehrmann}

\vspace*{1cm}
{\it DESY, Theory Group, D-22603 Hamburg, Germany}\\

\vspace*{2cm}

\end{center}

\begin{center}
{\bf Abstract}
\end{center}

The results of a recent calculation of the QCD corrections to 
the longitudinally polarized Drell-Yan process are summarized.
In addition to the parity conserving double spin asymmetry 
which is already accessible in the Drell-Yan production of lepton pairs 
at fixed target energies, we consider parity violating single spin
asymmetries appearing only in the production of massive vector bosons
at polarized hadron colliders. The prospects for extracting 
polarized quark distributions from asymmetries in vector boson
production at RHIC are briefly discussed.

\vspace{6.5cm}
\noindent
{\it Invited talk presented at the workshop ``Deep Inelastic Scattering
off Polarized Targets: Theory meets Experiment'', Zeuthen, Germany, 1997.}
\vfill

\setcounter{page}{0} 
\newpage

\begin{center}
{\LARGE\bf QCD Corrections to the}

\vspace*{2mm}
{\LARGE\bf Polarized Drell-Yan Process}

\vspace{1cm}
{\large T.~Gehrmann}

\vspace*{1cm}
{\it DESY, Theory Group, D-22603 Hamburg, Germany}\\

\vspace*{2cm}

\end{center}

\begin{abstract}
The results of a recent calculation of the QCD corrections to 
the longitudinally polarized Drell-Yan process are summarized.
In addition to the parity conserving double spin asymmetry 
which is already accessible in the Drell-Yan production of lepton pairs 
at fixed target energies, we consider parity violating single spin
asymmetries appearing only in the production of massive vector bosons
at polarized hadron colliders. The prospects for extracting 
polarized quark distributions from asymmetries in vector boson
production at RHIC are briefly discussed.
\end{abstract} 
\section{Introduction}

\vspace{1mm}
\noindent

The production of lepton pairs in hadron collisions, the Drell--Yan 
process~\cite{drellyan}, is one of the most powerful tools to probe the 
structure of hadrons. Its parton model interpretation is straightforward 
--~the process is induced by the annihilation of a quark--antiquark
pair into a virtual photon which subsequently decays into a lepton pair. 
The Drell--Yan process in proton--proton or proton--nucleus collisions
therefore provides a direct probe of the antiquark densities in protons 
and nuclei. Experimental data on the Drell--Yan process in unpolarized 
collisions are crucial to constrain the behaviour of the sea quark 
distributions at large $x$ and to determine the flavour structure of 
the light quark sea. It is therefore natural to expect that a measurement
of the Drell--Yan cross section in polarized hadron--hadron collisions
will yield vital information on the polarization of the quark sea 
in the nucleon, which is presently only poorly constrained~\cite{fits,gs,grsv}
from deep inelastic scattering data~\cite{disdata}. 

Apart from the invariant mass $M$ distribution, one usually studies the 
distribution of the lepton pairs as function of the Feynman parameter 
$x_F$ or of the hadron--hadron centre-of-mass rapidity $y$. The resulting 
distributions at fixed invariant mass can be directly related to the  
$x$-dependence of the parton distributions in beam and target. Moreover,
most fixed target experiments have only a limited kinematic coverage in
$x_F$ or $y$, such that only these distributions can be measured without 
extrapolation into experimentally inaccessible regions.

All calculations for Drell--Yan cross sections can be immediately 
applied to the production of vector bosons at hadron colliders, which 
is mediated by exactly the same partonic subprocesses. In the case of 
vector boson production with polarized beams, one  finds however one 
crucial difference to the Drell--Yan process, which can be viewed 
as the production of an off-shell photon of invariant mass $M$.
The parity 
violating couplings of the vector bosons give rise to 
non-vanishing single spin asymmetries, which are absent in the 
'classical' Drell-Yan process. These single spin asymmetries, measured 
in vector boson production with one polarized and 
one unpolarized hadron beam, allow for a study of the various polarized 
quark distributions. The single spin asymmetries can 
at lowest order be directly related~\cite{old} 
to the ratio of polarized and unpolarized quark distributions, and they 
are largely dominated by one particular quark (or antiquark) flavour in 
most regions of $y$. 

The double and single spin asymmetries in massive vector 
production will soon become experimentally accessible with the 
start of the spin physics programme at RHIC~\cite{rhic}. It will on 
the other hand be very difficult to study the Drell--Yan process 
at RHIC, since its cross section is already vanishingly small at 
RHIC energies ($\sqrt s = 500$~GeV). The ideal place to study this 
process would be e.g.~the proposed HERA-${\vec{{\rm N}}}$ 
experiment~\cite{heran} 
operating a fixed polarized nucleon target in the polarized HERA 
proton beam~\cite{herapol}.

The rather large QCD corrections to the unpolarized Drell--Yan cross 
section~\cite{unpol} suggest that a reliable interpretation of the 
Drell--Yan process in terms of partonic distribution functions is only 
possible if higher order corrections are taken into account. Following
closely the method of the unpolarized calculation~\cite{unpol}, we have 
calculated~\cite{dypol} 
the next-to-leading order corrections to the $x_F$- and $y$-distributions 
of lepton pairs produced in collisions of longitudinally polarized hadrons. 
This calculation has recently been extended~\cite{dynew} 
to single spin asymmetries in vector boson production at hadron colliders.

A fully consistent numerical 
study of spin asymmetries in the Drell--Yan process and 
in  vector boson production
 at next-to-leading
order was until now not possible, as the polarized 
parton distributions could only be determined at leading accuracy. With the 
recently calculated polarized two--loop splitting functions~\cite{nlosplit},
the polarized distributions can now be determined to next-to-leading order
from fits~\cite{fits,gs,grsv} 
to polarized structure function data. We will present some 
consistent next-to-leading order results for these asymmetries 
below.

\section{Double and single spin asymmetries at next-to-leading order}
\vspace{1mm}
\noindent
Truncated up to ${\cal O}(\alpha_s)$, 
the cross section for the double polarized Drell--Yan process 
receives contributions from the 
$q\bar q$--annihilation process at leading and next-to-leading order 
and the quark-gluon Compton scattering process. It can be 
expressed as:
\begin{eqnarray}
\frac{{\rm d}\Delta \sigma_{LL}}{{\rm d}M^2 {\rm d}x_F} & = & \frac{4\pi \alpha^2}{9 M^2 S} 
\sum_i e_i^2 \int_{x_1^0}^1 {\rm d}x_1 \int_{x_2^0}^1 {\rm d}x_2 \nonumber \\
& & \hspace{-0.3cm}
\times \Bigg\{\left[\frac{{\rm d}\Delta \hat{\sigma}_{q\bar{q}}^{(0)}}
{{\rm d}M^2 {\rm d}x_F} (x_1,x_2) +\frac{\alpha_s}{2\pi}
\frac{{\rm d}\Delta \hat{\sigma}_{q\bar{q}}^{(1)}}
{{\rm d}M^2 {\rm d}x_F}\left(x_1,x_2,\frac{M^2}{\mu_F^2}\right)\right]
\nonumber \\
& & \hspace{0.5cm}
\Big\{ \Delta q_i(x_1,\mu_F^2)\Delta\bar{q}_i(x_2,\mu_F^2) 
+\Delta\bar{q}_i(x_1,\mu_F^2)\Delta q_i(x_2,\mu_F^2) \Big\} \nonumber \\
& & \hspace{0.1cm} + \Bigg[ \frac{\alpha_s}{2\pi}
\frac{{\rm d}\Delta \hat{\sigma}_{qg}^{(1)}} {{\rm d}M^2 {\rm d}x_F} \left(x_1,x_2,
\frac{M^2}{\mu_F^2}\right)\nonumber \\ 
& & \hspace{0.5cm}
\Delta G(x_1,\mu_F^2) 
\left\{ 
\Delta q_i(x_2,\mu_F^2) + 
\Delta \bar{q}_i (x_2,\mu_F^2) \right\} 
+ (1 \leftrightarrow 2)\Bigg] \Bigg\},
\label{eq:xfmaster}
\end{eqnarray}
where analytic expressions for the parton level cross sections in the 
$\overline{{\rm MS}}$--scheme 
are listed in~\cite{dypol}. The 
expression for the $y$-distribution of the Drell--Yan pairs takes a 
similar form. Integration over $x_F$ yields the Drell--Yan 
mass distribution which agrees with earlier results~\cite{check}.
\begin{figure}[t!] 
\begin{center}
~ \epsfig{file=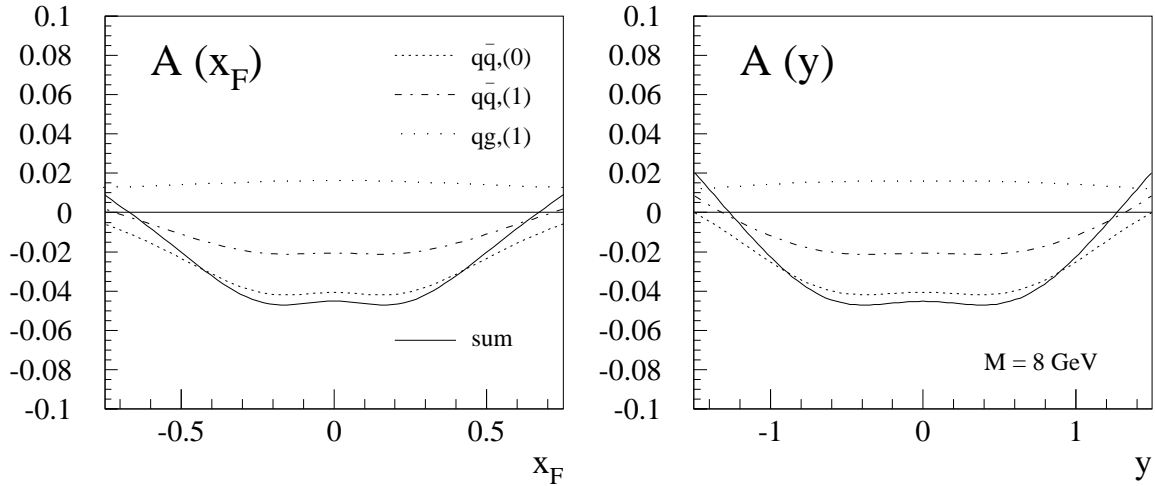,angle=-90,width=15.5cm}
\caption{Contributions of the individual parton level subprocesses 
to the double polarized Drell--Yan cross section (see text). }
\label{fig1}
\end{center}
\end{figure}

The numerical importance of these different contributions 
is illustrated in Figure~\ref{fig1}, 
which shows the ratio between polarized and 
unpolarized Drell--Yan cross sections for the collision of a proton 
beam ($E_p=820$~GeV) on a fixed proton target~\cite{heran}. 
All curves are obtained with the polarized GS(A) parton densities~\cite{gs}
and are shown for $M=8$~GeV. The polarized 
subprocess contributions are normalized to the {\it full} unpolarized 
cross section at next-to-leading order. The relative magnitude of the 
individual contributions is similar to the unpolarized Drell--Yan 
process~\cite{unpol}.
The ${\cal O}(\alpha_s)$ correction to the $q\bar q$--annihilation process 
enhances significantly the lowest order prediction while the quark--gluon 
Compton process contributes with a sign opposite to the annihilation process. 
However, the relative magnitude of annihilation and Compton process depends 
on the magnitude of the gluon distribution at large $x$, which is 
completely undetermined at present~\cite{gs,grsv}. This uncertainty 
prevents the sensible prediction of a
$K$--factor between the cross sections at leading and
next-to-leading order. 

The theoretical uncertainty inherent to an extraction of the polarized 
quark distribution functions from future data on the polarized Drell--Yan 
process can be estimated by varying the mass factorization scale $\mu_F$ 
which is used in the evaluation of the asymmetry. It is found~\cite{dypol}
that the absolute value of the asymmetry changes by less than 0.01 if 
$\mu_F$ is varied between 0.5 $M$ and 2 $M$. 
This 
uncertainty has to be compared to the discrepancy between predictions 
for the asymmetry obtained with different parameterizations of the 
polarized parton distributions~\cite{gs,grsv}, which lie in the 
band $A=-0.05\ldots +0.05$. 

\begin{figure}[t!] 
\begin{center}
~ \epsfig{file=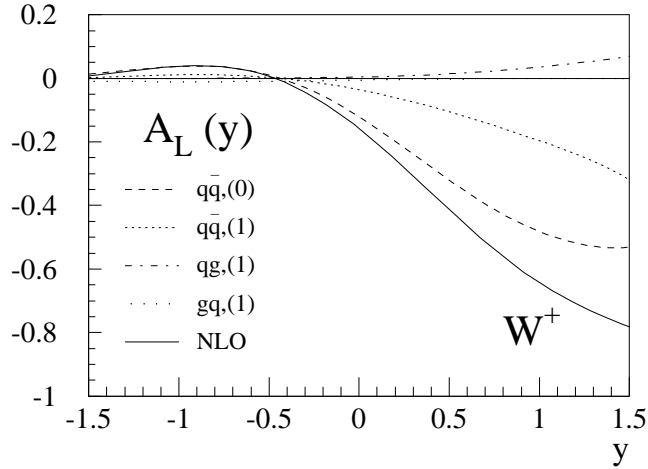,angle=-90,width=8.5cm}
\caption{Contributions of the individual parton level subprocesses 
to the single polarized $W^+$ production cross section at RHIC
(see text). }
\label{fig2}
\end{center}
\end{figure}

In the QCD corrections to 
single spin asymmetries~\cite{dynew,check2}, one has to discriminate 
two different contributions from the quark--gluon Compton scattering 
process: a gluon from the polarized hadron scattering of a quark from 
the unpolarized hadron (denoted by $gq$) and a gluon from the unpolarized 
hadron scattering off a quark from the polarized hadron (denoted by $qg$). 
The calculation of these QCD corrections is very similar to the 
corresponding calculation for the double spin asymmetries and will be 
presented in~\cite{dynew}. Figure~\ref{fig2} illustrates the magnitude of 
the individual partonic subprocesses to the single polarized $W^+$ 
production cross section in proton-proton collisions at RHIC 
($\sqrt s = 500$~GeV). As above, the contributions are normalized to
the full unpolarized cross section at next-to-leading order. The polarized 
proton is moving in the $+y$ direction, such that $A_L(y>0)$ is 
dominated by $\Delta u(x,Q^2)$ in the range $x \gsim 0.2$, 
while $A_L(y<0)$ mainly reflects $\Delta \bar d(x,Q^2)$
in the  $x \lsim 0.2$~\cite{old}. 

Figure~\ref{fig2}, obtained with the polarized GS(A) parton 
distributions for $W^+$ production at RHIC,
illustrates the sizable  impact of the 
next-to-leading order corrections on the single polarized cross section. 
Like in the case of the double polarized cross section, one finds an 
enhancement of the cross section due to corrections to the annihilation 
process, which is partly compensated by the Compton scattering 
process contributing with opposite sign. It is found that the 
contribution from the $gq$ Compton scattering process becomes 
relevant only for $A_L(y<0)$, while being completely negligible in 
$A_L(y>0)$. Consequently, an extraction of the large $x$ behaviour 
of the polarized quark distributions at next-to-leading order
 from a measurement of $A_L(y>0)$
at RHIC will not suffer from the uncertainty on 
the polarized gluon distribution, since only the $qg$ Compton scattering 
process (with the known unpolarized gluon distribution) contributes. 
The determination of the polarized antiquark distributions from $A_{LL}(y)$ or
$A_L(y<0)$ on the other hand requires always the knowledge of the polarized 
gluon distribution for a sensible prediction of the next-to-leading order 
corrections. 

Variation of the mass factorization scale $\mu_F$
allows again for an estimate of the theoretical uncertainty on the 
single spin asymmetries 
due to unknown higher order corrections. It is found that this uncertainty
is below 0.01 for the whole range of $y$, thus indicating the 
perturbative stability of the next-to-leading order calculation for
this asymmetry. Even smaller values are obtained for 
the double spin asymmetry $A_{LL}(y)$ in 
vector boson production at RHIC.

\begin{figure}[t!] 
\begin{center}
~ \epsfig{file=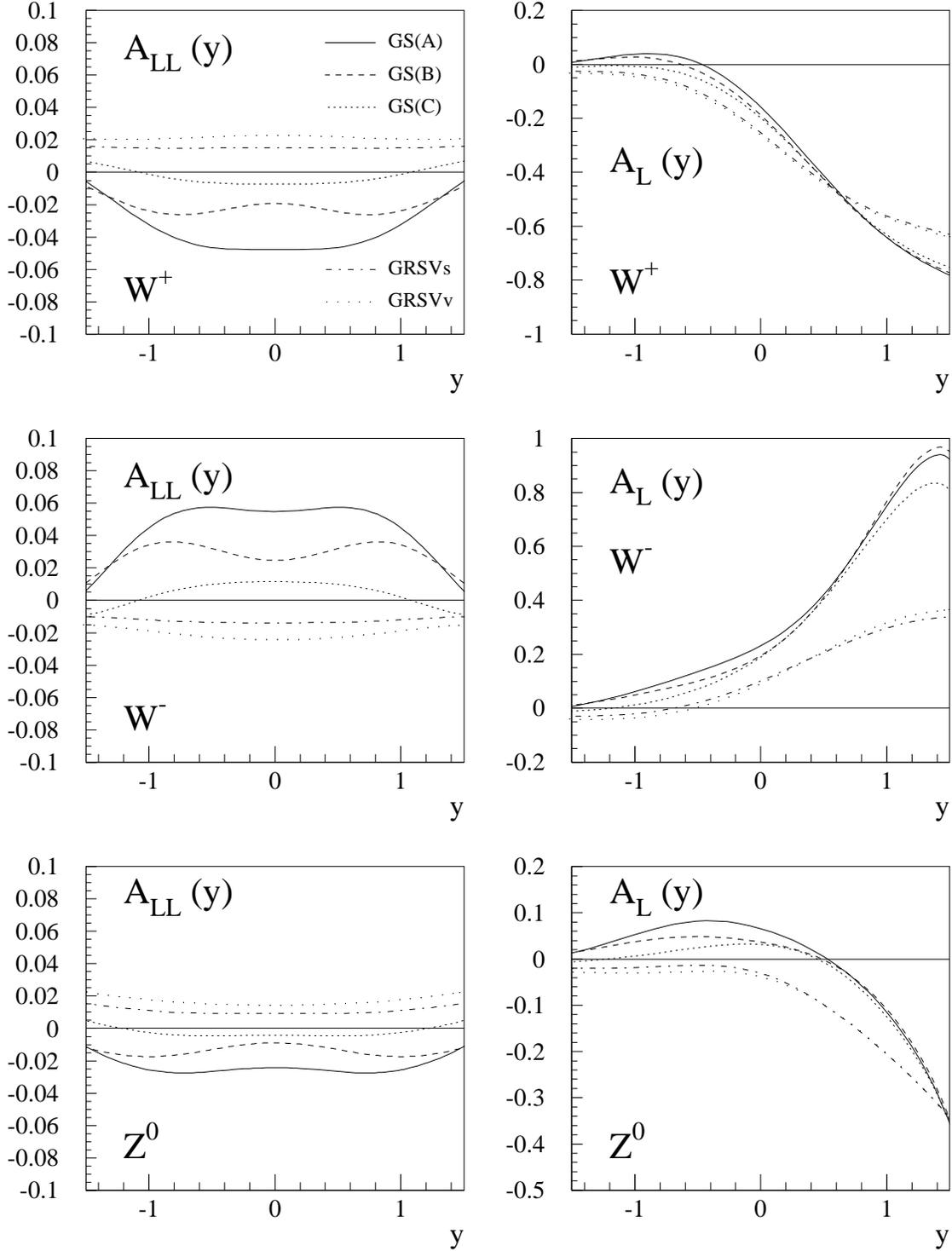,width=15cm}
\caption{Single and double spin asymmetries in vector boson production 
at RHIC for different parameterizations of the polarized parton distributions
at next-to-leading order.}
\label{fig3}
\end{center}
\end{figure}

The double and single spin asymmetries in vector boson production at
RHIC are evaluated for different up-to-date parameterizations~\cite{gs,grsv}
of the 
polarized parton distribution functions in Figure~\ref{fig3}. It can clearly 
be seen that these parameterizations yield significantly different 
predictions for all asymmetries, thus reflecting the range of 
quark polarizations in the nucleon still allowed by present polarized 
deep inelastic scattering data. 

The single spin asymmetry $A_L(y>0)$ 
is dominated by contributions from valence quarks at large $x$, which 
are already
reasonably well constrained. However, in the case of $W^-$ production,
related to $\Delta d(x,Q^2)$, a discrepancy 
of 
about a factor of 2 between different parameterizations can still 
be seen in this region. A measurement of this asymmetry at RHIC will
thus provide valuable supplementary information on the polarized 
valence quarks, although they are believed to be already
well constrained 
from deep inelastic scattering data.

The situation is even more drastic for $A_L(y<0)$ and $A_{LL}(y)$. 
These asymmetries are dominated by contributions from polarized 
sea quarks, which are only weakly constrained by present deep 
inelastic scattering data. This is clearly reflected in Figure~3: 
on the basis of the present parameterizations, it is not even possible 
to predict the sign of these asymmetries, neither their magnitude. 
Measuring these asymmetries at RHIC will clearly provide the first 
direct information on the shape and the flavour 
structure of the polarized light quark sea.

The discrepancies between different parameterizations are far larger 
than the theoretical error on all asymmetries estimated above. It is 
therefore apparent that a measurement at RHIC will enable an unambiguous 
discrimination between the parameterizations, which is not affected 
by theoretical uncertainties. However, a proper extraction of the
polarized sea quark distributions at next-to-leading order
from $A_L(y<0)$ and $A_{LL}(y)$ will require some knowledge on the 
behaviour of the polarized gluon distribution at large $x$, which 
enters in the determination of the $K$-factor.

\section{Summary}
\vspace{1mm}
\noindent

In summary, the complete ${\cal O}(\alpha_s)$ 
corrections to the $x_F$- and $y$-dependence of the longitudinally polarized  
Drell--Yan cross section have been derived~\cite{dypol}. These 
results have been recently extended~\cite{dynew} to  double and 
single  spin asymmetries in vector boson production in collisions of 
longitudinally polarized hadrons. These corrections are quantitatively 
similar to the corrections in the unpolarized case and hence sizable even
at collider energies. They
enable a consistent next-to-leading order 
determination of the polarization of quarks in the 
nucleon from future measurements of lepton pair production at fixed 
target energies or from massive vector boson production at RHIC. 

A reliable quantification of the next-to-leading order effects in the 
polarized Drell--Yan process requires at least an approximate knowledge 
of the polarized gluon distribution $\Delta G(x,Q^2)$. The lack 
of information on this quantity affects in particular the asymmetries 
$A_{LL}(y)$ and $A_L(y<0)$, which both probe the sea quark 
polarization in the nucleon. The asymmetry $A_L(y>0)$, probing mainly
the quark polarizations at large $x$, is on the other hand completely
insensitive to the polarized gluon distribution. 

Finally, we have illustrated the discriminative power of a measurement 
of asymmetries in vector boson production at RHIC between different 
up-to-date parameterizations~\cite{gs,grsv}
 of the polarized parton distribution functions. 
It was especially shown that the scale uncertainty on these asymmetries 
at next-to-leading order is far smaller than the deviation between the 
different parameterizations.

\end{document}